\documentclass{ws-rv975x65}
\usepackage{ws-rv-van}             
\makeindex

\usepackage{graphicx}
\usepackage{amssymb}
\usepackage{amsmath}

\def\fref#1{Fig.~\ref{#1}}

\def\lb{\label}

\newcount\bozza \bozza=1
\ifnum\bozza=1
\newdimen\shift \shift=-2truecm
\def\lb#1{%
{\label{#1}\rlap{\kern\shift{$\scriptstyle#1$}}}}
\else\def\lb#1{\label{#1}} \fi

\begin{document}

\chapter{Some experimental tests of Tomonaga-Luttinger liquids}

\author[T. Giamarchi]{Thierry Giamarchi$^{1}$}
\address{$^{1}$DPMC- MaNEP University of Geneva, \\ 24 Quai Ernest-Ansermet CH-1211
Gen\`eve 4, Switzerland}

\begin{abstract}
The Tomonaga-Luttinger-Liquid (TLL) has been the cornerstone
of our understanding of the properties of one dimensional systems. This universal
set of properties plays in one dimension, the same role than Fermi liquid plays
for the higher dimensional metals. I will give in these notes an overview of
some of the experimental tests that were made to probe such TLL physics. In particular
I will detail some of the recent experiments that were made in spin systems and
which provided remarkable quantitative tests of the TLL physics.
\end{abstract}

\body

\section{Introduction}

Interactions lead in one dimension to a physics radically different than for the higher dimensional counterparts \cite{giamarchi_book_1d}. Indeed if in high dimensions it is possible for excitations resembling individual particles (such as the famous Landau quasiparticle excitations) \cite{landau_fermiliquid_theory_microscopics,Nozieres_book}
to exist, this is not the case in one dimension where interactions will clearly turn any excitation into a collective one.

As a result a new type of physics exists, that can be described by the universal concepts of Tomonaga-Luttinger liquids (TLL) \cite{haldane_effective_harmonic_fluid_approach,giamarchi_book_1d}.
This theory describing the physical properties of most of the one dimensional interacting systems (bosons, fermions, spins) has been first introduced as a perturbative expansion around idealized dirac dispersion. It then has been proven to be a universal concept, playing a similar key role in one dimension than the Fermi liquid played in higher dimensions.

Given the importance and universality of the TLL concept, and more generally of the unusual one dimensional physics, many experiments have
sought to probe such a physics. I will describe in this brief review some ``morceaux choisis'' of the hunt for TLL. I will first briefly review the salient features of one dimensional quantum systems.
I will then give a few examples of the experimental tests, and discuss in details some
of the most recent achievements that allowed for the first time for a \emph{quantitative} test of such TLL physics. Of course this review cannot pretend to be exhaustive and much more details on the vast body of experiments in the context of one dimensional physics can be found at other places in the literature (see e.g. \cite{giamarchi_book_1d,cazalilla_rpm_bosons} and references therein). Nevertheless it should give an idea of the state of the art in the experimental search for TLL.

\section{Some basics of TLL}

Let me recall in this section some of the properties of the TLL that will be useful in connection with the experiments.
The goal is of course not to do a full review of the TLL. The reader wanting to get the full details on TLL is referred to \cite{giamarchi_book_1d}. Two main properties are crucial in a TLL and we examine them separately

\subsection{Power laws and universality}

One of the most visible properties of the TLL is the fact that all correlation functions behave as power laws with some non-universal
exponents. For example, the density-density correlations for a system of quantum particles of density $\rho_0$ is given by
\begin{equation} \label{eq:powerll}
 \langle \delta_\rho(x,\tau) \delta\rho(0) \rangle = \frac1{r^2} + A_2 \cos(2\pi\rho_0 x) \left(\frac1{r}\right)^{2K} + A_4 \cos(4\pi\rho_0 x) \left(\frac1{r}\right)^{8K} + \cdots
\end{equation}
where $x$ is the spatial position, $\tau$ the imaginary time, $\delta\rho(r) = \rho(r) - \rho_0$, and $r = \sqrt{x^2 + (u \tau)^2}$. The $A_n$ are non-universal amplitudes that depend on the model, the value of the interactions etc.. All the powerlaw behavior depends on two numbers, nicknamed the Luttinger liquid parameters: $u$ which is the velocity of density excitations, and a dimensionless parameter $K$. Both $u$ and $K$ depend on the model and the value of the interactions. There are various ways that these parameters can be computed \cite{giamarchi_book_1d}. The formula (\ref{eq:powerll}) presents several remarkable features.

The first one is of course the powerlaw behavior. Normally we associate this to a critical state, such as the one that occurs exactly at
the critical temperature in a phase transition. One can physically understand why one dimensional quantum systems present such critical behavior by remarking that quantum fluctuations preclude, even at $T=0$ the breaking of a continuous symmetry. The system thus cannot really
order and is poised at the brink of order, thus behaving in a critical way. This powerlaw behavior is one of the hallmark of the TLL. It shows in all the correlation functions. It has many consequences that extend beyond the $T=0$ results. Since the theory is conformaly invariant one can deduce from the powerlaw (\ref{eq:powerll}) behaviors at finite temperature or for finite size systems.

The second, and very often overlooked remarkable fact, is the universality of the exponents. The various correlation functions may have different exponents, but they are \emph{all} functions of the \emph{sole} parameter $K$, which also controls the thermodynamics of the system. It means that although the exponents themselves are non-universal the relations between the exponents are. This is one of the strongest properties of a TLL, and rests on the fact that the fixed point Hamiltonian that describes the low energy properties of the system is
\begin{equation}
 H = \frac1{2\pi} \int dx (u K) (\pi \Pi(x))^2 + \frac{u}{K} (\nabla_x \phi(x))^2
\end{equation}
where $\phi$ and $\Pi$ are conjugate variables.
This universality is of course what makes the TLL description so useful. It allows to go beyond the approximations that are usually inherent in models such as Hubbard, etc. for which one has to make a caricature of the interactions. With the TLL description one can simply take the
value of $K$ from one of the experimental value, and then be sure that one has a faithful description of the properties of the system. In that sense the TLL plays exactly the same role than the Fermi liquid plays in higher dimensions.

\subsection{Fractionalization of excitations}

Another crucial property of one dimensional interacting system is the possibility for the excitations to fractionalize. In the high dimensional world we are used to the fact that if we make an excitation with some quantum numbers that correspond to the \emph{minimum}
possible quantum numbers (e.g. one adds or removes one electron, such as in photoemission, or flips one spin 1/2 thus creating a $S=1$
magnon excitation) then these quantum numbers are stable. The excitation can scatter, acquire a lifetime etc. but cannot be viewed as a composite object of \emph{smaller} quantum number excitations. The only exception (except for very bizarre models) are the Laughlin
quasiparticles in the quantum hall effect, which can carry a charge smaller than the one of the electron.

\begin{figure}
 \begin{center}
  \includegraphics[width=0.9\columnwidth]{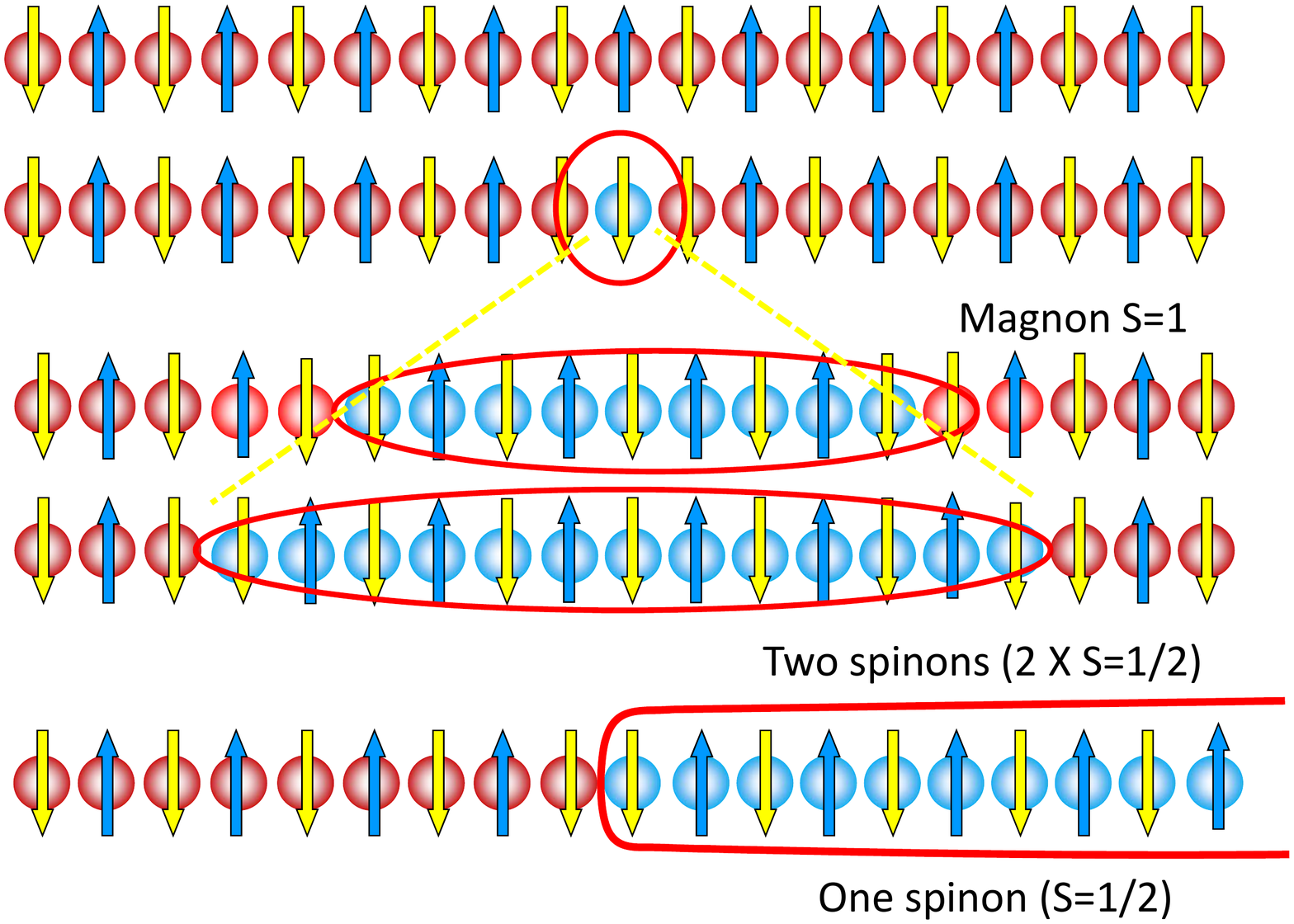}
 \end{center}
  \caption{\label{fig:magnon_spinon} In high dimensions the minimal spin excitation on an antiferromagnetic chain (top two lines) would
  be a magnon, in which a single spin is flipped. This is an $S=1$ excitation (2nd line). In one dimension the magnon
  is not an individual excitation, and a magnon decomposes into two $S=1/2$ excitations, the spinons (third and fourth line).
  Such excitations are soliton like objects with a string of spin flips on the right or the left of the position of the soliton (denoted by the blue circles), see bottom line for one spinon.}
\end{figure}

\begin{figure}
 \begin{center}
  \includegraphics[width=0.9\columnwidth]{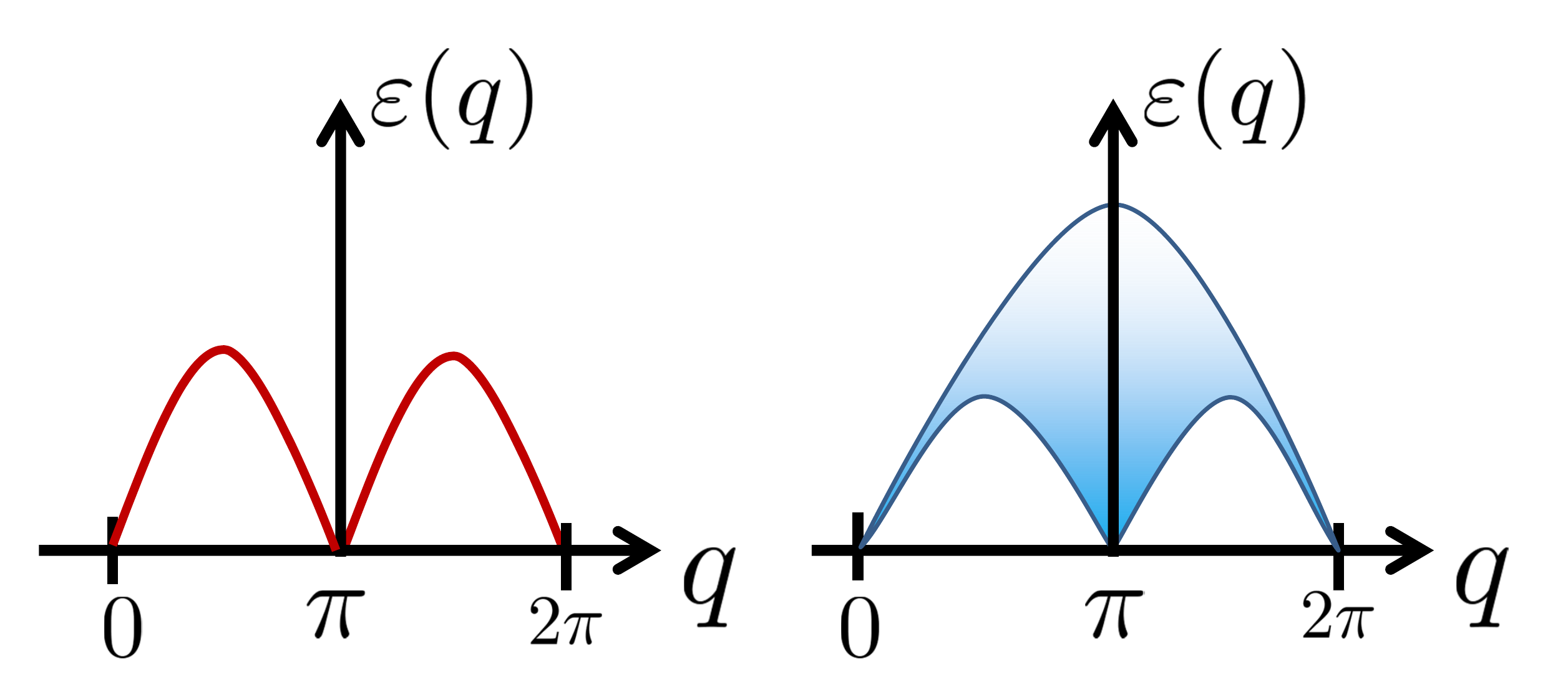}
 \end{center}
  \caption{\label{fig:spin_charge} (left) In high dimension a magnon excitation behaves as a particle, i.e. has
  a good relation between the momentum of the excitation $q$ and its energy $\epsilon(q)$. (right) In one dimension the magnon decomposes
  into two spinons (see text). As a consequence, there is now a continuum of excitations, the des Cloiseaux-Pearson spectrum.
  Such continuum can be clearly distinguished in neutrons scattering experiments.}
\end{figure}

In one dimension fractionalization is the rule rather than the exception. The very reason is that because of the one dimensional nature of the system it is possible to make soliton excitations, something that would be much more difficult in higher dimensions.
In \fref{fig:magnon_spinon} we show these two types of excitations. The spinon are the good excitations in one dimension.
Therefore for spinons there is a good relation between the momentum of the excitation $q$ and the energy of the excitation $E = J \cos(q)$.
This is not the case for the magnon excitation, which is made of two spinon excitations. As a result instead of a well defined relation between the momentum $q$ of the magnon and its energy $E$ there is a continuum since
\begin{equation}
 q = q_1 + q_2 \quad,\quad E = J\cos(q_1) + J\cos(q_2)
\end{equation}
where $q_i \in [-\pi/2,\pi/2]$ since spinons move by two lattice spacings, and one can spread the total momentum $q$ between the two spinon in any way compatible
with the above constraints. The leads to the spectrum shown in \fref{fig:spin_charge}.

This fractionalization of the excitations is even more spectacular in the case of an electron. In the presence of spin and charge, the good excitations are solitons of charge and solitons of spins. Hence the electron split in two excitations, the holon carrying charge but no spin
and the spinon which has spin but no charge \cite{giamarchi_book_1d}.

Note that the notion of fractionalization is more general than the TLL in itself. It is more resting on the 1D nature and note confined to the low energy properties of the system. Of course for several systems such property is also appearing in the low energy properties since terms
that couple the spin and charge excitations do in general exists (such as band curvature, impurities etc.). For more details on fractionalization we refer the reader to \cite{giamarchi_book_1d}.

\section{Power laws and such}

Let us first look at experiments that are probing the powerlaw behavior of a TLL. Of course it is out of the question to even attempt to make an exhaustive review here of all the experiments. The reader interested in that will find many experiments discussed in \cite{giamarchi_book_1d} and for bosons and spins in the recent review \cite{cazalilla_rpm_bosons}.

\begin{figure}
 \begin{center}
  \includegraphics[width=0.9\columnwidth]{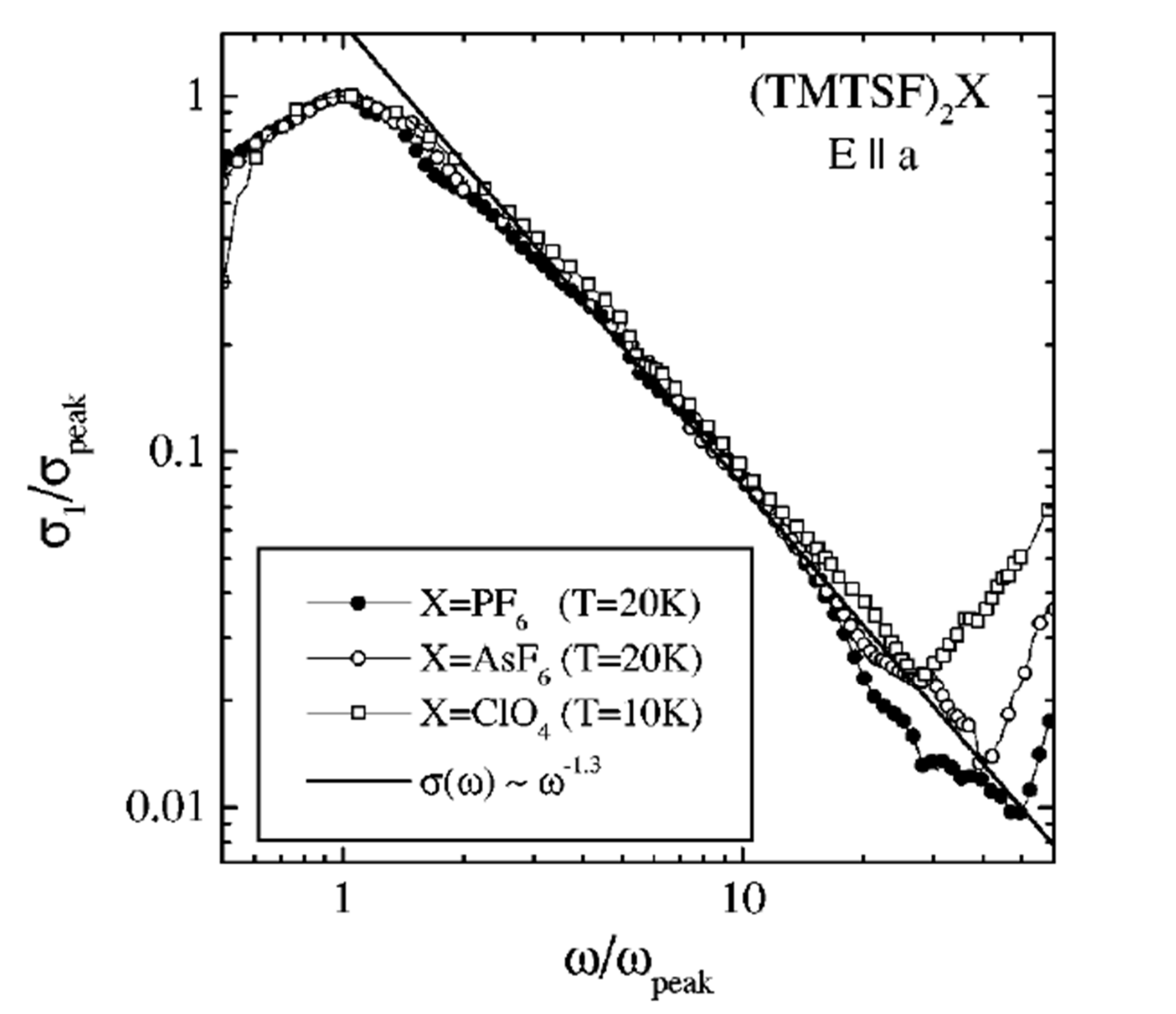}
 \end{center}
  \caption{\label{fig:organics} Optical conductivity of various organic conductors-superconductors, normalized to the observed
  optical gap. From TLL theory a power-law behavior is expected above the optical gap in excellent agreement with the observed spectra.
  This constituted the first observation of the TLL behavior in an electronic system. From \cite{schwartz_electrodynamics}.}
\end{figure}

The first experimental observation of powerlaws in a TLL for an electronic system, was done in the organic conductors \cite{lebed_book_1d}. These materials
are very anisotropic systems where stacks of organic molecules provide very different electronic overlaps for the band. Their bandstructure
can be well represented by a tight-binding model \cite{jerome_review_chemrev,bourbonnais_review_book_lebed}
\begin{equation}
 H = - \sum_{\langle i,j \rangle, \sigma} t_{ij} c^\dagger_{i\sigma} c_{j\sigma}
\end{equation}
with hopping along the three axis directions $t_a \sim 3000K$, $t_b \sim 300K$, $t_c \sim 20K$, where $a$ is the organic chain direction
and $b,c$ the two perpendicular directions. These systems have a commensurate filling and thus a tendency to become Mott insulators.
Because of the commensurate potential provided by the scattering of the electrons on the lattice, the high frequency conductivity should behave as a power law of the form \cite{giamarchi_umklapp_1d,giamarchi_mott_shortrev}
\begin{equation}
 \sigma(\omega) \propto \omega^{5 - 4n^2 K_\rho}
\end{equation}
where $K_\rho$ is the TLL exponents for the charge excitations, and $n$ the order of commensurability with the lattice ($n=1$ for half filling, $n=2$ for quarter filling etc.). Comparison of the above formula with measurements on the organic conductors is shown in \fref{fig:organics}.

The experiments clearly show over one decade in frequency the power law behavior. Additional experiments in such a class of materials, both on the temperature dependence of the conductivity \cite{jerome_review_chemrev,bourbonnais_review_book_lebed}, or on the transverse transport
\cite{henderson_transverse_optics_organics,dressel_transport_tmtsf,pashkin_pressure_optics_asf6,pashkin_pressure_temperature_optics_asf6}
have been found in agreement with TLL behavior. Similar properties were observed in other compounds \cite{lee_optics_chains_ybco}.
I refer the reader to the literature for more details on those points.

Shortly after the observation of the TLL power law in the organics conductors, transport in the nanotubes was investigated in a similar manner \cite{bockrath_luttinger_nanotubes}. In that case the weak contact between leads and the nanotube ensured that one was performing a tunneling experiment. Such experiments showed power law behavior of the conductivity that was again in nice agreement with the expected behavior in a TLL. Variation of such a power law was observed depending on whether the tunneling occurred at the edge or in the middle of the tube, or if an impurity was producing additional backscattering within the nanotube \cite{bockrath_defects_cnt_qd}. I refer the reader to the literature for more details on such class of experiments.

Tunneling inside a TLL can also be probed in bulk material by doing a photoemission experiment. The interpretation of such experiments is often delicate given the possibility of surface artefacts. Several powerlaw behavior have nevertheless been observed \cite{grioni_review_arpes_q1d,denlinger_arpes_LiMoO}. Some powerlaw has been
seen in photoemission in the organic conductors. However the ranger of energy over which such powerlaw was seen casts doubts on the fact that all of it is due to TLL physics, even if the extracted TLL parameter $K_\rho$ would be in a good agreement with the other measurements on this compound \cite{vescoli_photoemission_tmtsf}. Other one dimensional systems have also shown power law behavior under photoemission. One of the most interesting is provided by the purple bronze compound, where power laws in the density of states well consistent with the expected TLL behavior have been observed both
in photoemission and STM \cite{wang06_arpes_limo6o17,hager_stm_purple_bronze}. The high energy behavior is well consistent with a TLL behavior and strong interactions, while there are still some puzzles in this compound to be explained for the low energy behavior \cite{wang_purple_arpes,giamarchi_physics_purple}.

Another bread class of systems in which power laws were investigated was provided by realization of the TLL using edge states such as in the quantum hall effect \cite{chang_review_edges}. Indeed the two chiral states existing at the edge of a Hall systems can be viewed as the right and left movers of a standard TLL. It can be shown that the TLL exponent $K$ is related to the fractional Hall plateau of the conductance of the system \cite{giamarchi_book_1d}. Backscattering between the edge can thus be analyzed in terms of backscattering in a TLL and such experiments, although posing some questions on whether quasiparticles or electrons tunnel, have provided evidence of the powerlaw scaling expected from a TLL. More details on that class of experiments can be found in the literature.

Last but not least cold atoms have provided remarkable realizations of interacting quantum system \cite{bloch_cold_lattice_review}, in which the TLL behavior can be in principle quantitatively tested. Experiments on interacting bosons have successfully probed the so called Tonks limit in which hard core bosons are found to behave as spinless fermions, showing the common behavior of the one dimensional quantum fluids that is covered by the
TLL universality class \cite{cazalilla_rpm_bosons}. However in experiments in optical lattices the ability to probe for the TLL power laws is impaired by the fact that such systems are inhomogeneous because of the trap, an $r^2$ varying chemical potential. Since the TLL depends on the density and interaction the inhomogeneity obscures the simple power law behavior. This is obvious in the Tonks limit for which the TLL behavior would have given a single particle correlation decaying as $1/\sqrt{x}$ (corresponding to $K=1$ the limit for hard core bosons or free spinless fermions), and for which the averaging of many tubes lead to quite different and much less clear behavior. However experiments on atom chips do not suffer
from such inhomogeneity and indeed interferences between two bosonic tubes have shown behavior well compatible with TLL \cite{hofferberth_interferences_atomchip_LL}. However in these
last experiments the interactions are quite weak for the moment, leading to extremely large TLL parameters ($K \sim 42$) making it
very hard to unambiguously distinguish TLL theory for simpler ones such as a quasi-condensate or a time dependent Gross-Pitaevskii theory \cite{pitaevskii_becbook}.
Fermionic systems are for the moment at too high a temperature for the TLL properties to be probed in them.
It is clear that the field is rapidly progressing and that many of the objections of today will soon be overcome, in particular with
the advent of local probes, that allow to avoid the inhomogeneity problems. In particular recently predicted \cite{berg_haldane_cold_bosons} non-local order parameters,
characteristic of 1D behavior have been observed \cite{endres_string_mott_cold}. There is thus little doubt that cold atoms will offer remarkable and controlled realizations of the TLL in the future.

\section{Spin charge separation}

Another crucial aspect of the 1D behavior, namely the fractionalization of excitations \cite{giamarchi_book_1d} has also been observed.

\begin{figure}
 \begin{center}
  \includegraphics[width=0.9\columnwidth]{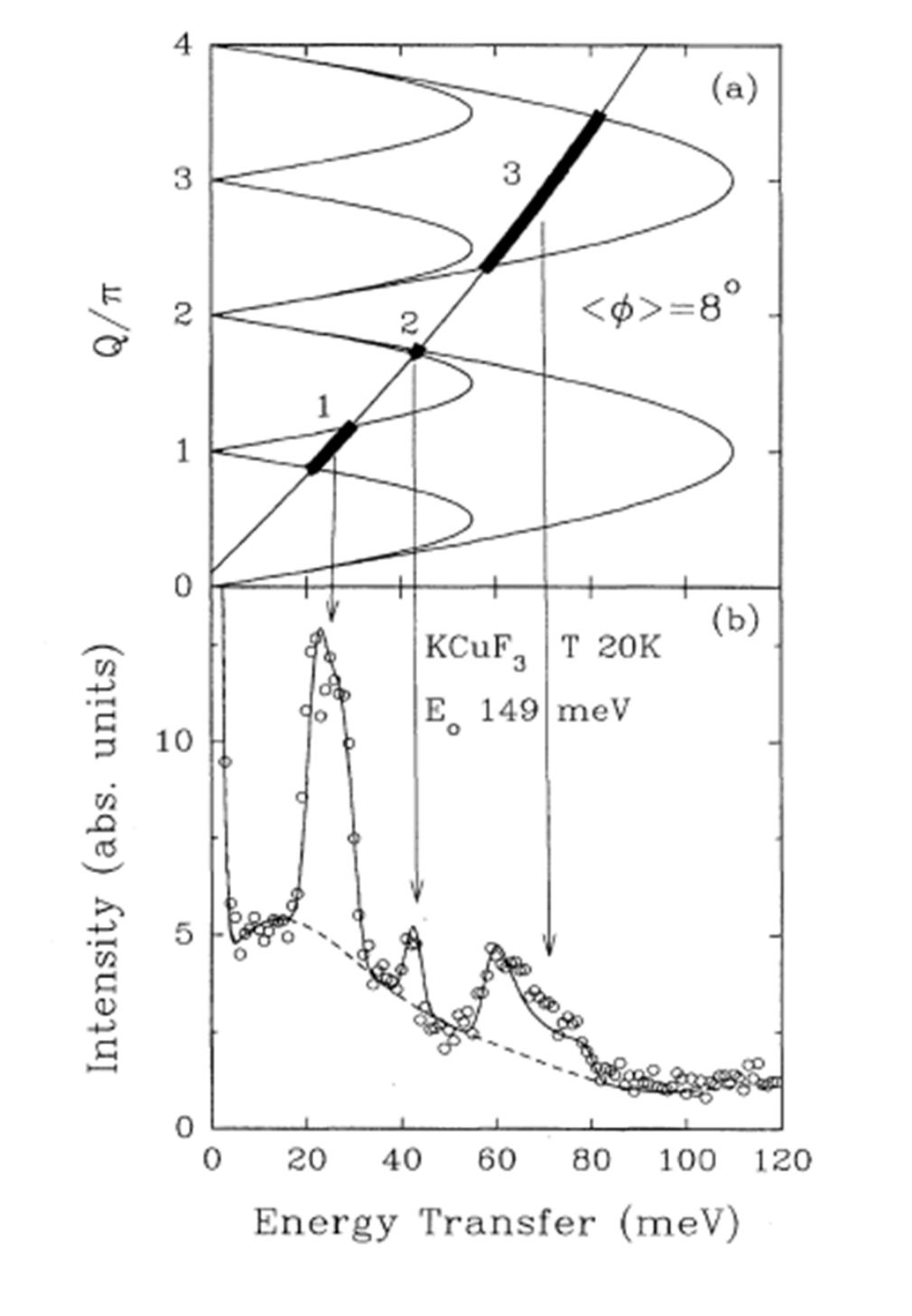}
 \end{center}
  \caption{\label{fig:spinchains} (a) The line represents the scan in energy $E$ and momentum $q$
  which is done in a neutron scattering experiment. If one neglects the matrix elements specific to the
  spin-spin correlation function measured, one can expect intensity within the continuum described in \fref{fig:spin_charge}.
  (b) The corresponding intensity in a neutron scattering experiment on the spin chain compound KCUF$_3$.
  The intensity is in excellent agreement with the prediction of the fractionalization of the magnon excitation inherent
  to a one-dimensional situation. Note that to get the precise lineshape and not just the existence of the response,
  one would need the matrix elements in addition to the spectrum of excitations. From \cite{tennant_kcuf_1d}.}
\end{figure}

For such features the system of choice has proven to the quantum spin systems. Indeed in such systems the spectrum of excitations can
be probed in a very clear way by neutrons scattering experiments. Although of course the neutrons scattering probes the full spin spins correlation functions one can view it as a gross determination of the spectrum if one uses the Lehmann representation of the spin-spin spectral function measured in a neutron scattering experiment
\begin{equation}
 S(\omega,q) = \sum_{\nu} |\langle \nu | S^-_q | 0 \rangle|^2 \delta(\omega + E_0 - E_\nu)
\end{equation}
where $|0\rangle$ is the ground state of the system of energy $E_0$, $|\nu\rangle$ the exact eigenstates of energy $E_n$ and the operator
$S^-_q$ flips a spin with momentum $q$. Ignoring as a first approximation the matrix elements $\langle \nu | S^-_q | 0 \rangle$ one can
expect intensity only within the continuum of excitation. One can thus well distinguish between a quasiparticle like excitation, and a continuum. Some of the first experimental results for spin chains are shown in \fref{fig:spinchains}.
The so-called Des Cloiseaux-Pierson spectrum is clearly identified, showing the fractionalization of the magnon excitations into two spinon ones in one dimension. of course since then more refined experiments have been performed in a variety of systems and the continuum of excitations clearly confirmed. I refer the reader to \cite{stone_spectrum_spinchain,thielemann_spectrum_spinladder,bouillot_dynamics_ladder_DMRG_long}
for more details on these more recent experiments.

Beyond spin chains, probing the fractionalization in spinfull systems, namely the spin charge separation is not an obvious task.
One possible route is provided by photoemission, in which one would expect that because of the spin-charge separation the standard quasiparticle peak would split in two singularities, one at $\omega = u_\rho k$ and one at $\omega = u_\sigma k$ reflecting the existence
of the holon and the spinon. Although some photoemission experiments claim to have observed the two excitations and their dispersion, the observation remains difficult and very often controversial. To the best of my knowledge the only experiment that unambiguously observed the spin-charge separation was the one involving the tunneling between two quantum
wires \cite{auslaender_quantumwire_tunneling,tserkovnyak_quantumwire_tunneling}.
Control provided in such experiments by a gate voltage
(which controls the difference of energy between the two wires) and a magnetic field (allowing to control the difference of momentum)
allows to tunnel from one of the Fermi points in one wire to one in another wire. The spin charge separation of the electron into two independent modes allows to see two different branches in the tunneling spectrum. Interference using the finite length of the wire allow
to confirm that there are indeed two different modes with two different velocities.

Cold atoms, given the level of control should be the ultimate systems in which to test for the spin-charge separation. No experiment
has been realized so far but theoretical proposals \cite{kleine_2velocities_bosons}
have been put forward on how to test for it. The rapid progress in probing
should give measurements of this feature in a not too distant future.

\section{Quantitative tests of TLL}

As is obvious when looking at the previous selected sets of experiments for TLL physics, and of course the many other realizations (see
\cite{giamarchi_book_1d,cazalilla_rpm_bosons} and the various chapters of the present book), the situation can be both considered as satisfactory and also quite frustrating. On the positive points:
\begin{enumerate}
 \item There is a growing number of one dimensional materials, both in the field of condensed matter and in the rapidly evolving field of cold atomic systems.
 \item Power laws have been observed in a variety of materials such as organics, nanotubes and cold atoms, and by a large variety of
 probes (optical conductivity, d.c. transport, photoemission, stm etc.)
 \item The effects of fractionalization has been evidenced, both for spinless (spectrum in spin chains), and spinful systems (quantum wires, photoemission).
\end{enumerate}
The TLL physics is thus definitely on the map and applicable to a growing number of materials.
On the other hand although some of the properties of TLL have been unambiguously seen, some problems still remain:
\begin{enumerate}
 \item In most of the tests for the electronic systems, the exponent is an adjustable parameter. This is due to the fact that for electronic systems it is extremely hard to determine from ab initio calculations the value of the TLL parameter $K$ since the interaction is a screened long range coulombic interaction.
 \item In most of the systems the universality of the TLL description, i.e. the fact that all the exponents depend on a single value of the parameter $K$ has not been tested. This comes from the fact that it is often difficult to do more than one type of measurement on a given compound. Some exception is provided by the organics. For more on that point we refer the reader to \cite{bourbonnais_review_book_lebed}.
 \item In most of the experiments there is no control parameter, such as the band filling or the control of the interactions. In that respect it is of course possible to apply pressure to change the interactions, but the global effects are complicated to evaluate.
     In the cold atom context it is possible to vary the interactions or bandwidth in a controlled way but not on systems free of the harmonic trap, which clearly complicates the identification of Luttinger exponents.
\end{enumerate}

It is thus highly desirable to have \emph{quantitative} tests of the TLL and to be able to plug the above-mentioned holes.
Cold atoms are of course candidate of choice in that direction. They can easily realize one dimensional bosonic or fermionic systems.
The interactions are short range and very controlled. Unfortunately the presence of the harmonic trap and/or of many inequivalent tubes
makes the identification of TLL powerlaw behaviors extremely difficult since the exponents depends on the interaction and density of particles.
As discussed in the previous section, atom chips, which are very homogeneous, could provide a solution. For the moment however,
interaction is these systems is yet to weak to  unambiguously distinguish the TLL behavior from theories such as time dependent Gross-Pitaevskii physics.

It is thus suitable to have another type of systems in which such ``quantum simulation'' of TLL systems can be done. Quite remarkably such systems have been provided by quantum spin systems. I will thus introduce some aspects of these systems, and refer the reader to the published literature for more detailed references.

\subsection{Quantum ladders and spin systems as quantum simulators}

The drawback of being unable to precisely know the interactions in a condensed matter context can be circumvented in a Mott insulator.
In that case the charges being localized, only spin superexchange remains. The localized quantum magnets thus offer the advantage to have short range, interactions. If one deals with the standard Heisenberg of XXZ hamiltonian one has
\begin{equation} \label{eq:xxz}
 H = \frac{J_{XY}}2 \sum_i [S^+_i S^-_{i+1} + S^+_{i+1}S^-_i] + J_Z \sum_i S^z_i S^z_{i+1} - h \sum_i S^z_i
\end{equation}
using the well known Holstein-Primakov representation $S^+_i \to b^\dagger_i (-1)^i$ and $S^z = b^\dagger_i b_i - \frac12$,
were $b$ represent hard core bosons one can map the problem (\ref{eq:xxz}) to an itinerant problem
\begin{equation}
 H = - t \sum_i [b^\dagger_i b_{i+1} + {\rm h.c.}] + V \sum_i (n_i - \frac12)(n_{i+1}-\frac12) - h \sum_i (n_i - \frac12)
\end{equation}
where $t = J_{XY}/2$ and $V = J_Z$. This is of course a very well known result \cite{auerbach_book_spins}.
One thus sees that a spin system could be used as
a quantum simulator of hard core bosons with a nearest neighbor interaction. As mentioned one big advantage is that the interactions
between the particles are short range and a priori well known if the exchange are known. Another important advantage of such a system
is the use of the magnetic field as a chemical potential. One can control the ``number'' of particles by changing the magnetic field and measure the number by simply measuring the magnetization. That allows to realize the equivalent of a gate voltage for charged particles and
thus to study the effects of interactions on the whole range of band filling. The empty band would be the fully polarized down chain, while
a filled band would be a fully polarized up chain.

\begin{figure}
 \begin{center}
  \includegraphics[width=0.9\columnwidth]{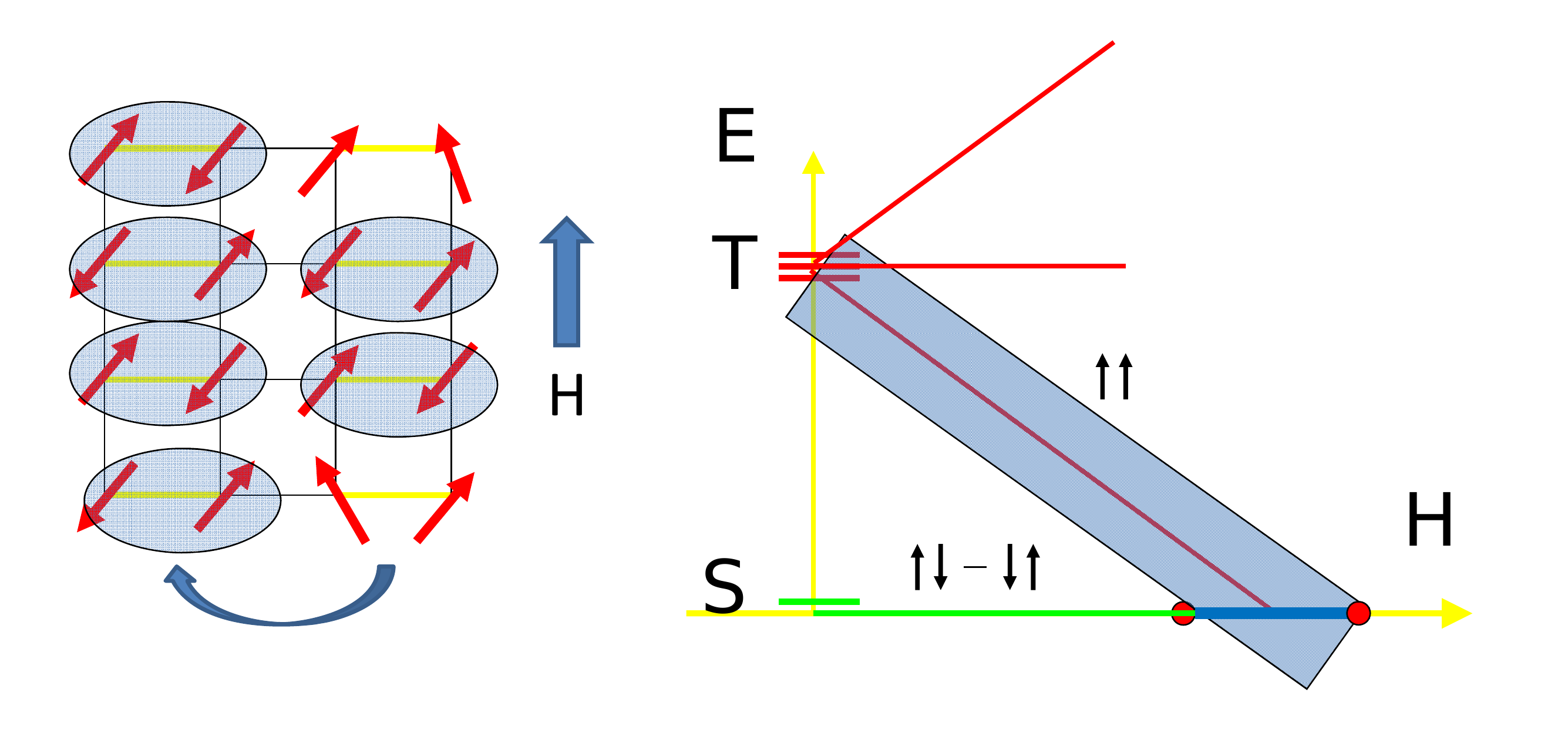}
 \end{center}
  \caption{\label{fig:ladders} Ladders under magnetic field. left) The ground state of dimer systems (the strong rungs $J_r$ denoted by he
  yellow lines) is made of singlets. This ground state is stable with respect to the weaker exchange $J$ denoted by black lines. Under the magnetic field $H$ some of the singlets are transformed into triplets, which can propagate from one rung to the next thanks to the exchange $J$. Using a mapping of the singlet to a vacuum of bosons, and the triplet to the presence of a hard core boson, this magnetic system realizes a tight binding model of interacting hard core bosons, whose density can be controlled by the magnetic field. right) the energy spectrum of the system on the left. The dispersion of the triplets (only the lower dispersion is shown) leads to the presence of two quantum critical points when $H$ is varied (denoted by red dots). At $H_{c1}$ the first triplet (``boson'') enters the system, while the band of triplets is full at $H_{c2}$.}
\end{figure}

Although these results were well known for a long time, using them directly was difficult. First, the typical spin exchanges are usually
in the range of several hundred Kelvin, which would mean impossible magnetic fields to polarize the chain and thus use the field as a control parameter. Second, since the two critical points, empty band and filled band correspond to a fully polarized system, there are usually parasitic couplings such as the dipolar forces that manifest themselves very strongly and thus spoil the nice mapping described above. So although the use of spin chains for this purpose is possible it is in general not easy. A system which is much more flexible is provided by dimers, either in a 3D, 2D or 1D (ladders) under magnetic field as shown in \fref{fig:ladders}.
I will discuss here mostly the case of the strong rung ladder, but similar arguments can be done for all cases. For strong rungs, in the absence of magnetic field, the ground state consists of singlets, while the magnetic field will favor triplets on the rung. One can use a similar mapping between the singlet and triplet states and the absence of presence of a boson on the rung. All commutation relations are obeyed. The ladders have the advantage that the two critical points are controlled by the two different energy scales $J_\perp$ and $J$.
In addition the singlet state is very robust so parasitic interactions are usually less effective. The mapping one the bosons can thus be implemented, and these systems used to ``simulate'' itinerant quantum bosons. In particular they were very fruitful to realize Bose-Einstain
condensation. I refer the reader to the review \cite{giamarchi_BEC_dimers_review} for more discussion and references on that point.

\begin{figure}
 \begin{center}
  \includegraphics[width=0.9\columnwidth]{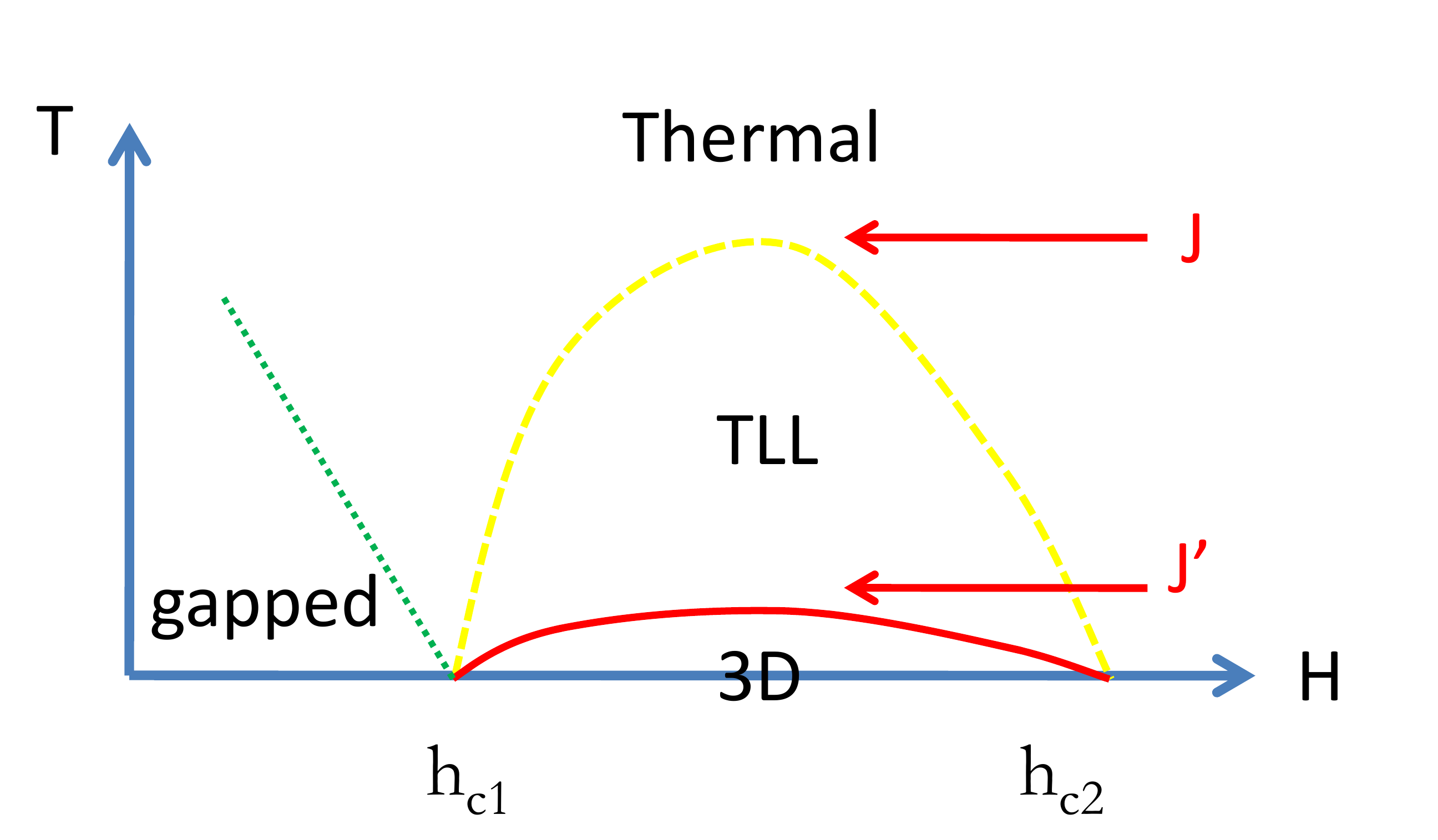}
 \end{center}
  \caption{\label{fig:phasediag} Phase diagram of Ladders under magnetic field. The phase below $H_{c1}$ or above $H_{c2}$ are gapped,
  (either the singlet phase or the fully polarized one). The gap (and hence the field $h_{c1}$ is of the order of the rung coupling $J_r$. Between the two critical fields the band of triplets is partly filled. There is a coherence scale (corresponding roughly to the distance between the bottom or the top of the band, below which the systems is described
  by TLL physics. The typical scale is given by the triplet dispersion along the legs of the ladders, of the order of the exchange $J$. Due to the weak coupling $J'$ between the ladders, there is an ordered antiferromagnetic phase in the XY plane, perpendicular to the magnetic
  field, that can take place.}
\end{figure}

In one dimension (i.e. for ladders) these systems can thus be used to simulate TLL. For one dimension the hard core bosons can be further mapped onto spinless fermions. The phase diagram of these systems is indicated in \fref{fig:phasediag}.

\subsection{Quantitative test of TLL - power laws}

Fortunately excellent experimental realization of such systems exist. One of them, that proved to be remarkably useful is provided
by the BPCB compound, an organic ladder. The exchange constant can be obtained directly from neutron scattering measurements. They can
also be directly extracted from a comparison of the measured magnetization and a computation using finite temperature Density Matrix
Renormalization group (DMRG) calculations. One can check, by comparing measured specific heat with the DMRG computed one that no important term
was forgotten in the Hamiltonian. Having the exchange constants allows for a \emph{quantitative} test of the TLL. The procedure is the following
\begin{enumerate}
 \item Compute from the Hamiltonian the TLL parameters. The general procedure is explained in \cite{giamarchi_book_1d}. Note that there
 is no adjustable parameter there. The values of $u$ and $K$ as a function of the magnetic field are totally determined. For an electronic system that would correspond going from an empty band to a full band.
 \item Compute several correlation functions of the system from the TLL theory and compare with the experimental ones. Note that now we have both a control parameter, which is the magnetic field, and the exponents are \emph{not} adjustable parameters but known quantities.
\end{enumerate}

This program can be implemented in practice for the BPCB. I will not give all the details here and refer the reader to \cite{klanjsek_nmr_ladder_luttinger,thielemann_neutron_ladder,bouillot_dynamics_ladder_DMRG_long} for
more information. In short we used for the various correlation functions: a) the NMR relaxation time $1/T_1$ which is related to the local spin-spin correlation function; b) because of the interladder coupling, as shown in \fref{fig:phasediag}, there is a transition towards a three dimensional ordered phase. The $T_c(h)$ can be computed from the TLL theory; c) the order parameter in the ordered phase. The list is of course not limitative, and other correlations can be computed as well, and compared to future experiments. An example is provided by the ESR resonance which was found to be in remarkable agreement \cite{furuya_ESR_BPCB}.

\begin{figure}
 \begin{center}
  \includegraphics[width=0.9\columnwidth]{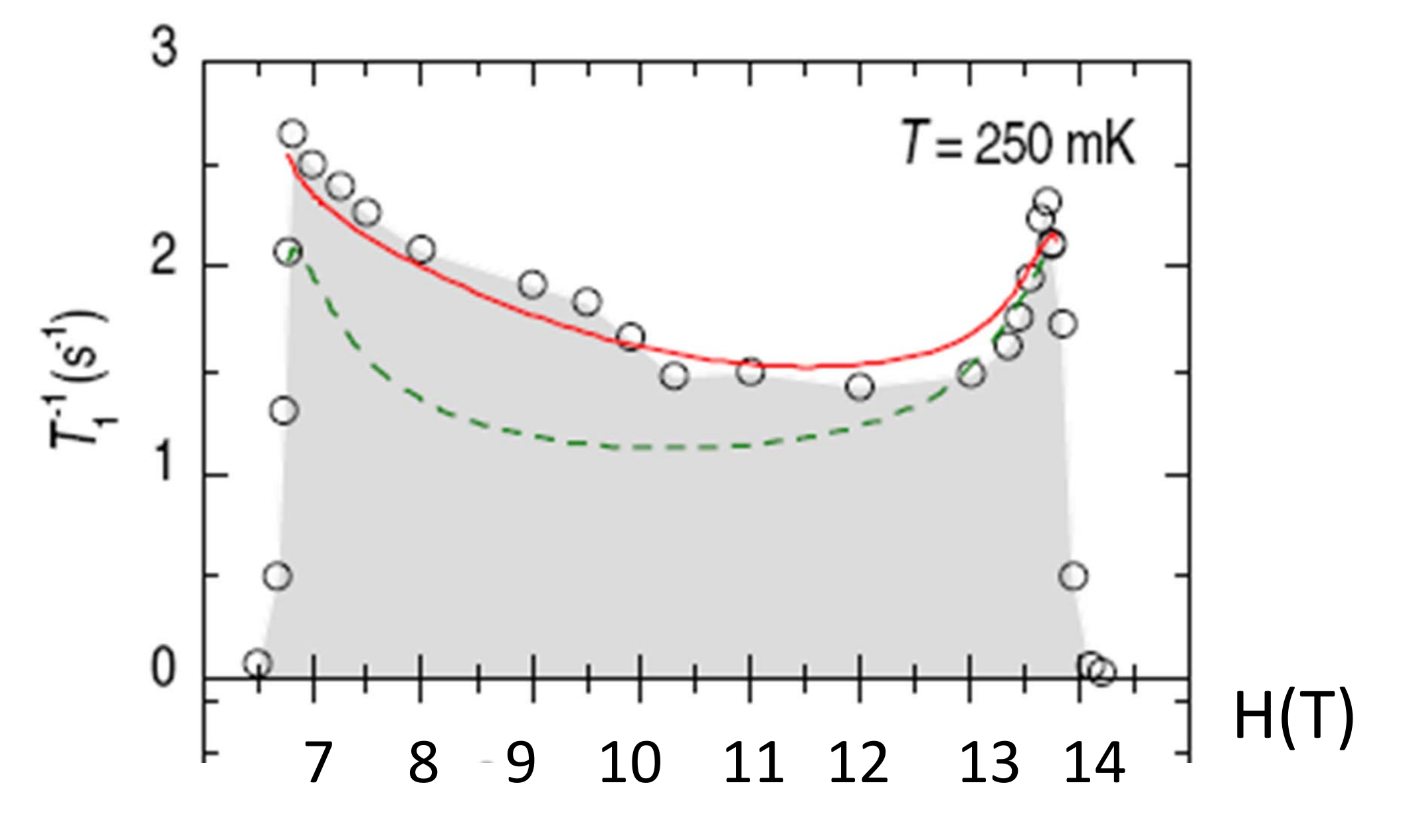}
 \end{center}
  \caption{\label{fig:nmr} NMR $1/T_1$ measured for the ladder compound BPCB (dots). The red line is a the theoretical calculation
  based on the TLL theory (see text). Note that the \emph{only} adjustable parameter here is the hyperfine coupling constant (which is in
  addition roughly known). Since the $1/T_1$ is obtained for the whole range of magnetic field one could only shift up and down
  the theoretical curve by changing this parameter but not the overall shape, which is totally fixed by the calculation of the TLL
  parameters, amplitudes and general formulas. The remarkable agreement with the experimental curve thus constitute the first
  \emph{quantitative} test of the TLL theory, where not only asymptotic behaviors with fitting exponents, but absolute the full dependence
  of observable quantities can be compared with experiments. After \cite{klanjsek_nmr_ladder_luttinger}.}
\end{figure}

As an example the NMR $1/T_1$ is given for a TLL by
\begin{equation} \label{eq:nmr-ll}
T_1^{-1} = \frac{1}{2} \gamma^2 \frac{\hbar}{k_B} A_\perp^2 \cdot
\cos\left(\frac{\pi}{4K}\right) B\left( \frac{1}{4K}, 1-\frac{1}{2K} \right) \cdot
\frac{2A_0^x}{u} \left( \frac{2\pi T}{u} \right)^{\frac{1}{2K}-1}
\end{equation}
where all the parameters in the above formula, except the hyperfine coupling, are fully known as a function
of the Hamiltonian and magnetic field ($u$, $K$, etc.). The \emph{full} functional dependence of the relaxation
time as a function of the temperature and magnetic field is thus known and can be quantitatively compared with the
experiment. The result for the experimentally measured $1/T_1$ is shown in \fref{fig:nmr}
The agreement is remarkable and provides a \emph{quantitative} test of the TLL predictions. It is worthwhile to note that the description is not only valid in some distant asymptotic regime, but very efficient
for practically all usable band filling. Such experiments demonstrates that the combination of analytic calculations based on the
TLL theory and the numerical calculation of the TLL parameters provides an extremely powerful theoretical tool. Of course for
the general experimental case, for which it is difficult to compute the TLL parameters ab-initio, one can still measure them in one
experiment and then use them for all the other experimental quantities.

\subsection{Full dynamical correlations}

The TLL theory allows to compute the dynamical correlations. However in many cases, one can be interested in such correlations for energies comparable to the bandwidth. This for example the case with neutron scattering experiments for which the neutron energy can be much larger than the exchange energy $J$. It is normally extremely difficult to obtain such correlation functions directly. Exact diagonalization leads usually to small systems, for which a direct comparison with experiments is difficult because of the finite size effects. Quantum Monte Carlo can compute dynamical correlations, but obtains the results in imaginary time. It is then necessary to perform the analytic continuation to
real time, using relatively uncontrolled methods such as maximum entropy. Fortunately in one dimension recent progress in the DMRG method
have made it possible to compute directly the correlation functions in real time, with an extremely good accuracy. I will not go in detail on the corresponding techniques, and relative merits of the various methods, and refer the reader to the literature for more details on that point \cite{cazalilla_rpm_bosons}.

\begin{figure}
 \begin{center}
  \includegraphics[width=0.9\columnwidth]{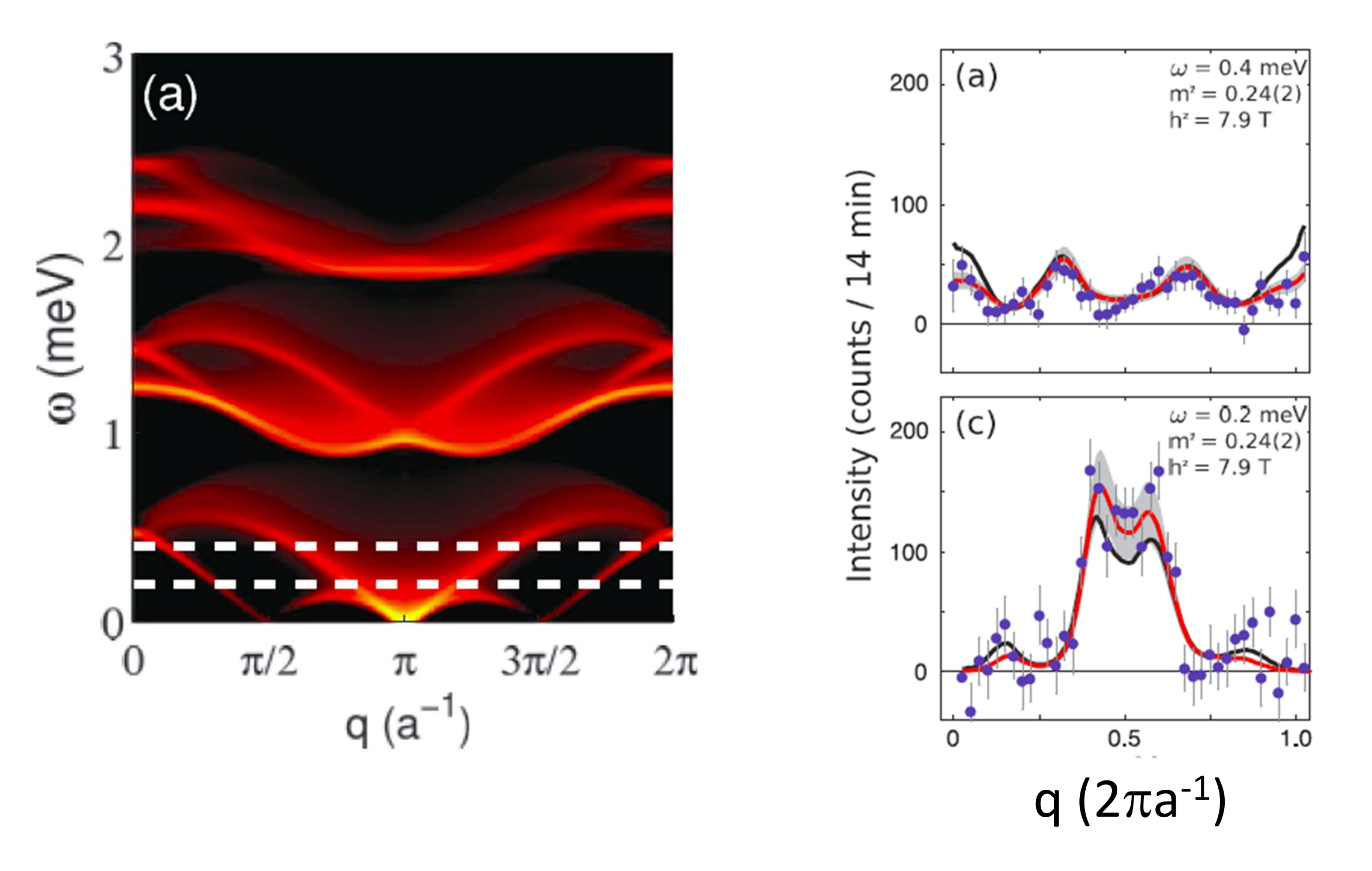}
 \end{center}
  \caption{\label{fig:bpcb-neutrons} (left) DMRG calculation of the dynamical spin-spin correlation for a spin ladder. The bottom part is the Des Cloiseaux-Pearson spectrum modified because of the incommensurability induced by the magnetic field. The higher parts of are reminiscicence of the other triplets, and are beyond the reach of the simple TLL description of the system. DMRG proves to be a very efficient way
  to obtain this part of the spectrum. (right) comparison with two scans of constant intensity ($E=0.2$ and $E=0.4$ meV) with neutron scattering experiments done one the compound BPCB. The agreement is excellent showing that we are now able to quantitatively compute
  such dynamical correlation functions. After \cite{bouillot_dynamics_ladder_DMRG_long}.}
\end{figure}

The DMRG can be successfully applied the case of the spin ladder systems, as shown in \fref{fig:bpcb-neutrons}
The resolution is about $J/20$ in a energy and a comparable figure for the momentum resolution, better than the existing experiments.
As a result one can clearly see the fractionalization of excitations and also predict additional features at high energy that can be directly tested experimentally and would be beyond the reach of an asymptotic theory. Of course the numerics fails at long time (low energy) but this is the regime where the analytic field theory (TLL) description is particularly efficient. The overlap between the two techniques is such that one does have a full description of the whole spectrum.

The possibility to compute such dynamical correlations with an accuracy that is beyond the experimental one opens very interesting perspectives. In particular the neutron spectrum is very rich and one can can thus use the comparison between the observed correlation functions and the essentially exact theoretical calculation to reconstruct the Hamiltonian of the system.
This program has been successfully carried out for the DIMPY compound \cite{schmidiger_dimpy_neutrons} for which it allowed to obtain very accurate values for the coupling constants. Such ab-initio reconstruction of the Hamiltonian is clearly something that will become a useful techniques in the future.

One can imagine using similar techniques to probe for fractionalization and spin-charge separation in other type of systems. For cold atomic gases proposals have been made but experiments still remain to be done.

\section{Conclusions and perspectives}

One dimensional systems have proven to be extremely rich. It has seen in the recent years extremely important progress both on the theory side and on the experimental one. On the theory side, the development of novel methods both analytical and specially numerical have put us in good position to compute reliable many of the 1D properties that were elusive before. Likewise the explosion of experimental systems in which one dimensional physics is at play allowed for extremely stringent test of such TLL physics.

As a result several of the key features of TLL physics, namely the power-law behavior of the correlation functions and the fractionalization of the excitations have been observed in several experimental systems. Recently \emph{quantitative} tests of the TLL physics could be performed in spin insulators. Such tests, which offered an extremely high level of control, both theoretically and experimentally fully confirmed the TLL predictions, not only for existence of power-laws and fractionalization but also as far as the universality of the dependence of the exponents in terms of the TLL parameters was concerned. Quite interestingly the TLL has proven not only to be accurate
in asymptotic regimes but also showed a remarkable accuracy at intermediate energies. More generally the combination of TLL analytical calculations and of numerical DMRG calculation of dynamical correlation functions have given access to the full physics of one dimensional
systems, for the whole energy range.

Such control on the theory side clearly opens the door to much more complex studies. In particular it allowed to do reconstruction of the
Hamiltonian directly from neutron scattering spectra. Such combination of techniques must now be used to tackle problems which before were
extremely hard, such as quansi-one dimensional systems and deconfinement, disordered or frustrated systems and of course out of equilibrium
physics. There is no doubt that important progress should come, making it extremely exciting times to look at one dimensional physics.

\section{Acknowledgements}

My own understanding of the one dimensional systems, and the several works described in these notes, have benefitted from so many people that it
would be impossible to mention them all there. The recent works on spin insulators result from collaborations with P. Bouillot and C. Kollath.
Several aspects of these works benefitted, on the theory side, from collaborations with S. Capponi, R. Chitra, R. Citro, S. Furuya, A. La\"uchli, E. Orignac, B. Normand, M. Oshikawa, D. Poilblanc, A. Tsvelik and, on the experimental side, close interactions and constant discussions with, in particular,
M. Klanjcek, M. Horvatic and C. Berthier for the NMR, B. Thielemann and C. R\"uegg for Neutrons on BPCB and D. Schmidiger and A. Zheludev for neutrons on DIMPY. This work was supported by the Swiss NSF under MaNEP and Division II.

%


\end{document}